\def\NOTES{1}

\documentclass[sigplan,10pt,nonacm]{acmart}

\usepackage{times}  
\usepackage{hyperref}
\usepackage{titlesec}
\usepackage{subcaption}
\usepackage{xspace}
\usepackage{enumitem}
\usepackage{xcolor}

\usepackage{placeins} 
\usepackage{comment}  

\usepackage{siunitx} 
\DeclareSIUnit{\nothing}{\relax}
\newcommand{\SIadj}[2]{\SI[number-unit-product={\text{-}}]{#1}{#2}}

\newcommand{\systemname}[0]{\textsc{Nostradamus}\xspace}

\newcommand{\smartparagraph}[1]{\vspace{.05in}\noindent\textbf{#1}}

\if \NOTES 1
 \newcommand{\mcnote}[1]{\textcolor{brown}{[marco: #1]}}
 \newcommand{\alnote}[1]{\textcolor{teal}{[alireza: #1]}}
 \newcommand{\hnote}[1]{\textcolor{violet}{[hamid: #1]}}
\else 
 \newcommand{\mcnote}[1]{}
 \newcommand{\alnote}[1]{}
 \newcommand{\hnote}[1]{}
\fi

\makeatletter
\newcommand{\DeclareLatinAbbrev}[2]{%
  \DeclareRobustCommand{#1}{%
    \@ifnextchar{.}{\textit{#2}}{%
      \@ifnextchar{,}{\textit{#2.}}{%
        \@ifnextchar{!}{\textit{#2.}}{%
          \@ifnextchar{?}{\textit{#2.}}{%
            \@ifnextchar{)}{\textit{#2.}}{%
              {\textit{#2.,\ }}}}}}}}%
}
\makeatother
\DeclareLatinAbbrev{\eg}{e.g}
\DeclareLatinAbbrev{\Eg}{E.g}
\DeclareLatinAbbrev{\ie}{i.e}
\DeclareLatinAbbrev{\Ie}{I.e}
\DeclareLatinAbbrev{\etc}{etc}
\DeclareLatinAbbrev{\etal}{et~al}

\def\first {$(i)$\xspace}
\def\Second{$(ii)$\xspace}
\def\Secondb{$(ii-b)$\xspace}
\def\third {$(iii)$\xspace}

\hypersetup{pdfstartview=FitH,pdfpagelayout=SinglePage}

\begin{document}


\title{
Just-in-Time Packet State Prefetching }

\author{Hamid Ghasemirahni}
\orcid{0000-0002-0034-5098}
\affiliation{%
  \institution{KTH Royal Institute of Technology}
  \city{Stockholm}
  \country{Sweden}
}

\author{Alireza Farshin}
\orcid{0000-0001-5083-4052}
\affiliation{%
  \institution{NVIDIA}
  \city{Stockholm}
  \country{Sweden}
}

\author{Dejan Kosti\'{c}}
\orcid{0000-0002-1256-1070}
\affiliation{%
  \institution{KTH Royal Institute of Technology}
  \city{Stockholm}
  \country{Sweden}
}

\author{Marco Chiesa}
\orcid{0000-0002-9675-9729}
\affiliation{%
  \institution{KTH Royal Institute of Technology}
  \city{Stockholm}
  \country{Sweden}
}

\renewcommand{\shortauthors}{Ghasemirahni, et al.}


\begin{abstract}
Could information about \textit{future} incoming packets be used to build more efficient CPU-based packet processors? Can  such information be obtained accurately?
This paper studies 
novel packet processing architectures that receive external ``hints'' about which packets are soon to arrive, thus enabling prefetching into fast cache memories of the state needed to process them, \textit{just-in-time} for the packets' arrival. We explore possible approaches to \first obtain such hints either from network devices or the end hosts in the communication and \Second use these hints to better utilize cache memories. We show that such information (if accurate) can improve packet processing throughput  by at least 50\%. 
\end{abstract}

\maketitle

\section{Introduction}
\label{sect:introduction}

The introduction of multi-100-Gbps links has made packet processing highly challenging on commodity hardware. To process packets at these high rates, without causing a large amount of queuing and performance degradation, recent works advocate: \first offloading computationally expensive operations to programmable network devices \& accelerators~\cite{cheetah-lb,hXDP,lb-offload,metron,pegasus} and/or \Second performing optimizations to maximize the benefits provided by the CPU's cache memories~\cite{redundant_elimination,farshin-packetmill,farshin-slice-aware,farshin-ddio,om-hamid,batchy,morpheus,ericsson-switch,RESQ-scheduling,flowmage}.
\vspace{0.2em}

While these efforts improve performance, stateful applications, relying on per-flow data structures, still struggle to achieve high throughput and low latency due to their large memory footprint (\ie large amount of per-flow state). Fast memories with low-latency access time, such as CPU caches or SRAMs on programmable switches, are essential for supporting high-throughput packet processing. Yet, their limited size makes it difficult to accommodate the large states required by stateful applications. 
On a CPU-based server, when the state needed to process a packet is not available in any of the cache memories, the CPU must fetch the state from the slow (yet large) DRAM memory and it puts the processing of a packet on ``stand-by'', an operation that is detrimental for throughput and latency (as shown in this paper). 
We observe increasing friction between computer architectures and network protocols: computer architectures increasingly require packets processed using the same state to be received in \textit{bursts} to sustain higher cache-hit ratio~\cite{om-hamid}, whereas networking protocols progressively \textit{pace} traffic for better network-level statistical multiplexing~\cite{bbr}. 
Existing high-throughput packet processors delay packet processing for tens or hundreds of microseconds to rebuild bursts of packets before processing them~\cite{om-hamid,gro-dpdk}, which are then transmitted as bursts, defying the purpose of pacing packets at the network level.

Our goal is to combine the best of two worlds by satisfying both networking and hardware requirements, thus achieving higher performance. Our work complements previous optimizations by facilitating stateful networking applications 
to benefit from cache memories, thereby achieving better performance at higher packet rates. We minimize memory accesses (or equivalently cache misses) by ensuring that the essential data structures required for packet processing are already available in the cache when packets arrive. In particular, we focus on a stateful network function, specifically an L4 load balancer, and try to \first investigate the impact of \emph{looking into the future and prefetching the per-flow states before a packet arrives} and \Second explore different possibilities \& challenges of building a system, called \systemname, which provides information regarding upcoming packets and enables \emph{just-in-time} prefetching. We exploit the prefetching capabilities supported by x86 processors to potentially increase performance \emph{without} imposing any additional cost.  
We focus on CPU-based packet processing, yet similar principles could apply to any system relying on a hierarchy of memories with different access-time latencies, \eg a programmable switch with an internal fast \& small SRAM and an external DRAM memory.

This work is still underway, but the aim is to spur discussion of effectively performing 
packet state prefetching to better exploit cache memories. Preliminary results show that carefully-timed packet state prefetching results in up to 50\% throughput improvement on a (stateful) L4 load balancer when dealing with a large number of flows. Our work primarily focuses on load balancers because they are a fundamental building block in the current data center architecture, the performance of which directly affects the Internet services latency; however, our results and takeaways could be applicable to other stateful 
networking applications.

\smartparagraph{Contributions.} 
We
    \first show the negative impact of statefulness on performance, as
    cache misses cause a
    3$\times$ throughput drop (\S\ref{sect:motivation:perf}),
\Second highlight the benefits of carefully-timed state prefetching (\S\ref{sect:idea}), and
\third discuss the challenges of building a just-in-time prefetcher (\S\ref{sect:approaches}) while exploring future directions to extend such a system (\S\ref{sect:usecases}). We plan to release our source code to facilitate the reproducibility of our experiments.

\section{Background and Motivation}
\label{sect:motivation}

Section \ref{subsect:lb-background} provides essential background on 
load balancers' implementation while Section \ref{sect:motivation:perf} shows the impact of the state size on their performance.

\subsection{Load Balancers}
\label{subsect:lb-background}

Load balancers are one of the most critical network functions deployed in today's data centers. They ensure incoming requests are distributed efficiently among backend servers based on a pre-defined policy (\eg round-robin) and the content of the packet header/payload. For instance, an L4 load balancer typically uses the network- and transport-layer headers  (\eg the TCP/IP 5-tuple) to dispatch the incoming packets, whereas an L7 load balancer may use the application content (\eg HTTP URL)~\cite{lb-offload}. 

When distributing packets to backend servers, it is essential to send the subsequent packets of the same flow to the same server to prevent connection dropping and daisy chaining~\cite{beamer}. To do so, many providers rely on stateful load balancers (rather than stateless ones) to guarantee per-connection consistency~\cite{cheetah-lb}. 

Stateful load balancers typically employ hashing data structures to ensure fast lookups (\eg Cuckoo hashing) or fast insertions (\eg chained hashing)~\cite{massimo-connection}. 
Cuckoo hashing~\cite{cuckoo-hash} is increasingly used in many networking applications (\eg the DPDK hash library uses the Cuckoo hash algorithm to resolve collisions).

\smartparagraph{Cuckoo hashing} is an open-addressing algorithm that employs multiple hash functions to assign each key to multiple locations (aka buckets) in order to resolve collisions. It guarantees worst-case $\mathcal{O}(1)$ lookup \& deletion times. In cases where the number of entries is lower than 50\% of the hash table's capacity, Cuckoo hashing also offers $\mathcal{O}(1)$ insertion times. While a very large hash table would guarantee fast insertion times, it may prevent the hash table entries from staying in (or requests to be served from) the cache, as the hardware prefetching mechanisms might prefetch sparse buckets \textit{and} evict useful information from the cache. Most Cuckoo-hashing implementations use only two hash functions, where the locations are referred to as primary and secondary buckets. Some extensions of Cuckoo hashing store multiple entries per bucket to achieve better performance by reducing the number of memory loads~\cite{cuckooExtension-2007balanced}. 
The DPDK implementation of Cuckoo hashing maintains two tables. The first table is an array of buckets that contains a signature of the key and an index to the second table, whereas the second table is an array of stored keys along with their associated data.
To minimize memory accesses for a lookup, DPDK stores 8 entries per bucket (this value is configurable); hence, each bucket can fit on a single cache line (\ie \SI{64}{\byte}) and the system is able to examine up to 8 entries with a single memory-fetch operation.

\subsection{Impact of Statefulness on Performance}
\label{sect:motivation:perf}

The goal of this section is to show the impact of statefulness on performance. More specifically, we hypothesize that increasing the application state size could eventually cause the states to \emph{not} fit in the cache, causing performance degradation. 
To verify our hypothesis and understand the impact of memory footprint on the performance of load balancers, we use a simple L4 load balancer implemented on top of DPDK-based FastClick~\cite{fastclick}. We intentionally rely on DPDK, a kernel-bypass framework, to be able to \emph{purely} consider the load balancer states and exclude other stateful operations done within the Linux network stack\footnote{Evaluating the impact of statefulness within the Linux kernel remains as our future work.}. Our load balancer implementation distributes flows among $K$ servers in a round-robin fashion. To do so, it chooses a destination IP address (\ie a server) when the first packet of each flow arrives and then keeps the chosen address in the $i^{th}$ index of an array. The value of $i$ is stored in a Cuckoo hash table, enabling the load balancer to retrieve the chosen address upon the arrival of other packets belonging to the same flow. 

\smartparagraph{Testbed.} Experiments were run on a testbed containing two commodity servers connected together via a 32\,$\times$\,\SIadj{100}{Gbps} Edgecore Networks DCS800 Wedge 100BF-32X switch equipped with an Intel\textsuperscript{\textregistered} Tofino\texttrademark  ASIC~\cite{tofino}. One server acts as a traffic generator, and the other is our Device Under Test (DUT) which runs a stateful load balancer.
The DUT is equipped with NVIDIA Mellanox ConnectX\textsuperscript{\textregistered}-5 NICs~\cite{connectx5} and Intel\textsuperscript{\textregistered} Xeon\textsuperscript{\textregistered} Gold 6246R CPUs @ \SI{3.40}{\giga\hertz} with  2$\times$\SIadj{32}{\kibi\byte} per-core L1 (instruction \& data) \& \SIadj{1}{\mebi\byte} per-core L2 caches and a \SIadj{36}{\mebi\byte}  shared 
Last Level Cache (LLC) with 11 cache ways.
The Tofino switch enables us to \first make clones of incoming packets in a known order to increase the offered load on the DUT and \Second write the \hbox{5-tuple} hash of current \& predicted packets into the packets. We plot median values with min/max error bars (though in many experiments the range is small and almost invisible).

\smartparagraph{Workload generation.}
We use Fastclick~\cite{fastclick} on the traffic generator to create synthetic traffic traces (with UDP flows) and inject them into the network.
We minimize the locality of packets in the generated trace files by keeping the packets of each flow far from each other\footnote{Real-world traffic may have a different inter-packet distance distribution; further evaluation remains as our future work.}.
Additionally, we make multiple clones of each packet coming from the generator to put a higher load on the DUT. We change the source UDP port number for each cloned packet to avoid creating a batch of packets belonging to the same flow.
For the sake of simplicity and enforcing predictability, we only change the UDP ports between different flows in our experiments. Moreover, we warm up the hash table with the relevant entries to ignore insertion overhead and ensure that all the state lookups are served at a constant time.

\begin{figure}[h!]
    \centering
    \includegraphics[width=1\linewidth]{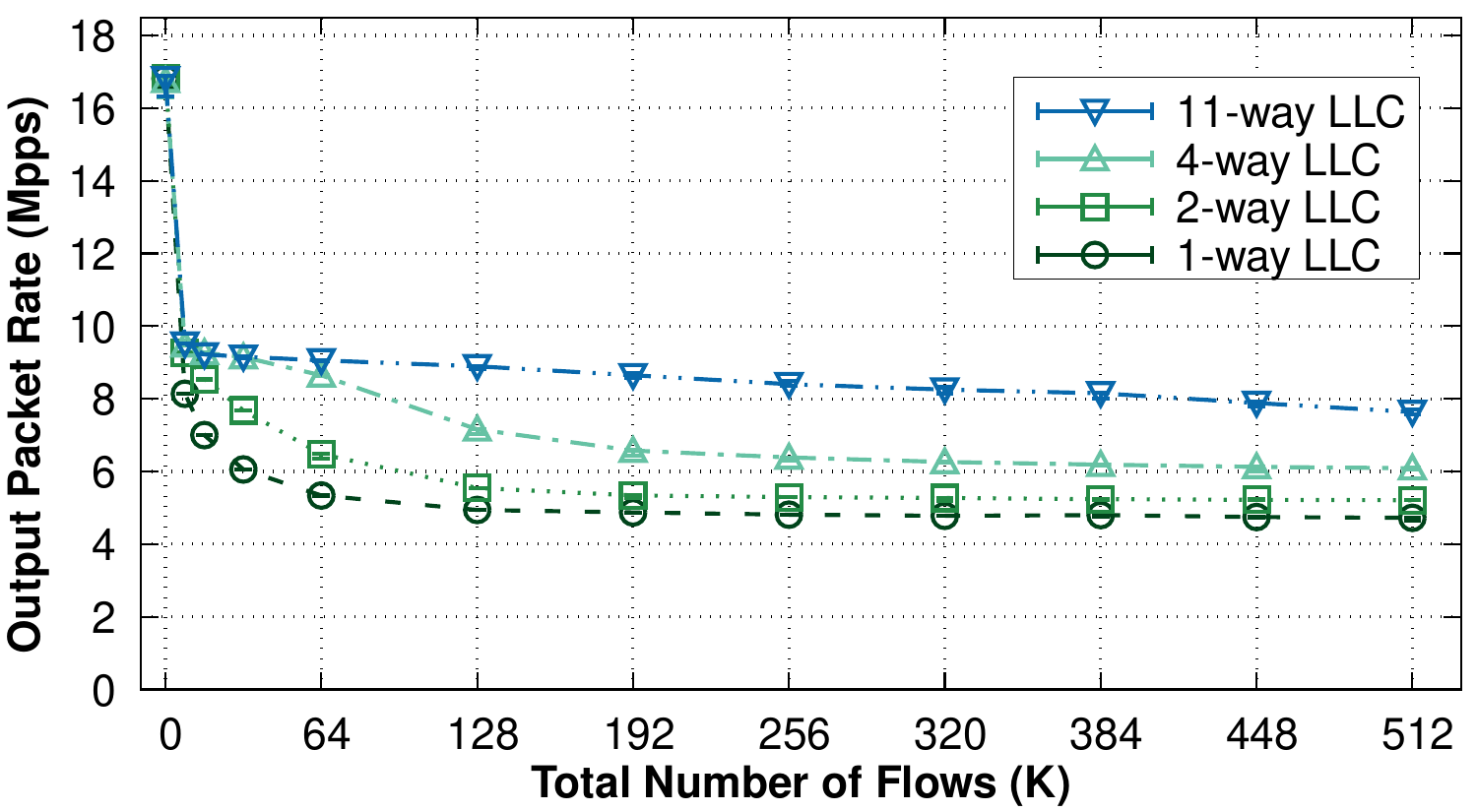}
    \caption{Increasing the number of flows causes an exponential decay in throughput of an L4 load balancer.}
    \label{fig:motivation:flow:pps}
\end{figure}

\smartparagraph{The performance of stateful NFs drops with increasing numbers of flows.} The first experiment measures the per-core throughput of a load balancer when the total number of flows increases. Intel's Cache Allocation Technology (CAT)~\cite{CAT} is utilized to change the size of the LLC allocated to the working core. 
Figure~\ref{fig:motivation:flow:pps} shows that the NF's throughput decays exponentially with an increasing number of flows. Reducing the size of the LLC causes the throughput to drop earlier, \ie with a smaller number of flows. For instance, when the working core is limited to use only 1 way (\ie \SI{3.5}{\mebi\byte}), throughput drops almost 70\% after exceeding \SI{64}{\kilo\nothing} flows. We expect a larger LLC quota to experience the same throughput drop but with a larger number of flows.


\begin{figure}[h!]
    \centering
    \includegraphics[width=1\linewidth]{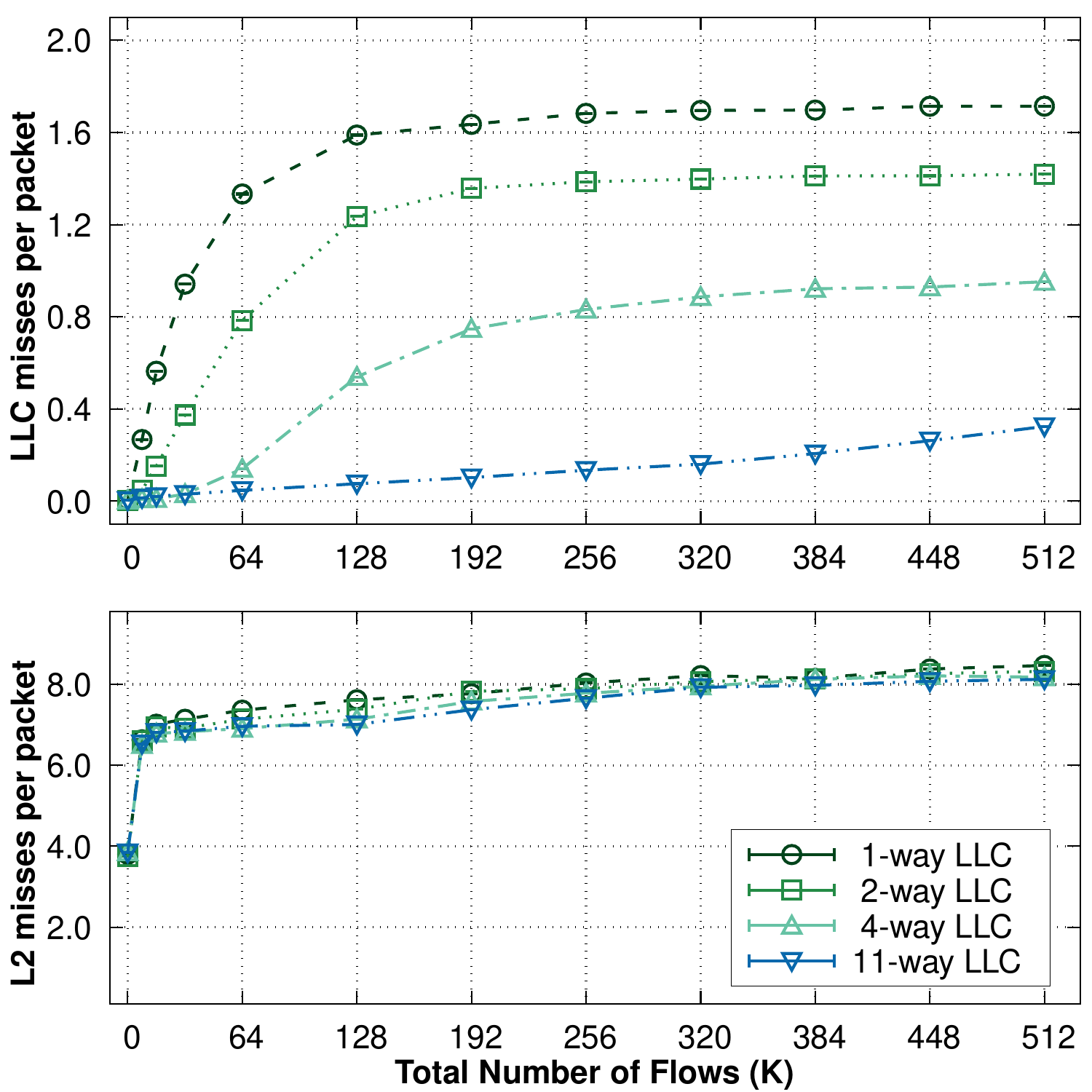}
    \caption{The average number of per-packet LLC misses increases with larger numbers of flows, which is inversely proportional to the throughput. The exponential increase in the number of per-packet L2 misses corresponds to the initial throughput drop.}
    \label{fig:motivation:flow:llc}
\end{figure}

To understand the reason behind the throughput drop, we measure the number of per-packet cache misses in each experiment. Figure~\ref{fig:motivation:flow:llc} illustrates that the drop in throughput is correlated with the increase in the number of per-packet LLC misses, which verifies our hypothesis that increasing the size of the state would prevent stateful network functions from benefitting efficiently from the cache memories. Additionally, the exponential increase in the number of per-packet L2 misses\footnote{We use \texttt{l2\_rqsts.miss} event to measure L2 misses, \ie similar trend as LLC loads in a non-inclusive cache hierarchy~\cite{scalable_family}.} justifies the sharp throughput drop after exceeding \SI{8}{\kilo\nothing} flows. Table insertions are \textit{not} the bottleneck since we have warmed up the table.

The rest of this paper will focus on the smallest LLC size to: \first emulate advanced applications with a larger memory footprint (to compensate for our simple low-overhead load balancer implementation), \Second consider real-world scenarios where multiple applications running on the same hardware have their own cache allotment to ensure security, privacy, and performance~\cite{contention-aware-justine,RESQ-scheduling}, and \third show the full potential of our solution when states do not fit into the cache. Investigating the impact of other cache allotments remains our future work. 
\section{
Prefetch the State in Advance}
\label{sect:idea}

The last section showed that increasing the state size prevents stateful network functions benefitting from the cache. To alleviate this, we exploit prefetching instructions (\eg \texttt{PREFETCHn}) offered by modern processors to issue memory load requests in advance and thus fetch data into the cache earlier than the actual packet processing time. This section investigates the impact of \emph{state prefetching}, \ie before packets arrive. Some highly optimized frameworks prefetch some data structures and/or some parts of the received packets at packet reception time, \eg FastClick~\cite{fastclick} prefetches the content of the packet and its metadata when receiving a batch of packets via DPDK, while FAJITA~\cite{fajita} and VPP~\cite{vpp-paper} utilize software prefetching to provide required state data for processing a batch of packets in stateful network functions.  In contrast to previous works, we use \textit{software prefetching mechanisms} to minimize the memory loads for \emph{future} batches of packets, which is complementary to the earlier efforts.

\subsection{Challenges and Solutions} 

While sounding appealing, performing 
prefetching requires addressing the following challenges:

\smartparagraph{What to prefetch?} It is important to find the trade-off between the benefits and overheads of prefetching, as sometimes prefetching the data may be very expensive due to many dependencies \& extra processing. For instance, DPDK-based hash tables store 8 hashes per bucket; therefore, loading the actual data stored in a bucket requires performing up to 8 comparison operations, which may cancel out the benefits of prefetching. Additionally, premature prefetching is known to be detrimental to performance~\cite{prefetching-journal}; prefetching a large amount of data may cause more harm than good, \ie this may evict the other useful data/code from the cache. We use large hash tables in our experiments, which increases the chance of finding data at the primary location of the key. Therefore, we only prefetch the primary bucket of the upcoming packet to avoid unnecessary prefetching. However, prefetching data from the secondary location could also be beneficial when the probability of finding data at the secondary location is higher (\eg when the hash tables are smaller).

\smartparagraph{How long to keep the data?} Intel's prefetching instructions 
make it possible to specify a temporal and spatial locality factor for data. Table~\ref{tab:prefetch} shows the specification of different x86 prefetching instructions~\cite{intel-prefetch}. Our experiments mainly use \texttt{prefetcht0} to maximize the state locality. The next section shows the impact of different locality factors. 

\begin{table}[h!]
\centering
\caption{Prefetching instructions in Intel processors.}
\label{tab:prefetch}
\resizebox{1\columnwidth}{!}{%
\begin{tabular}{|l||c c c|c|}
\hline
\multicolumn{1}{|c||}{\textbf{Instructions}} & \multicolumn{3}{c|}{\textbf{Cache Levels}} & \textbf{Early Replacement} \\ \hline \hline
\texttt{prefetcht0}                           & L1 & L2 & LLC                               & $-$                          \\ \hline
\texttt{prefetcht1}                               & $-$ & L2 & LLC                                   & $-$                          \\ \hline
\texttt{prefetcht2}                               & $-$ & L2& LLC                                   & $\times$                          \\ \hline
\texttt{prefetchnta}                              & $-$ & L2 & $-$                                     & $\times$ \\ \hline                        
\end{tabular}}
\end{table}

\smartparagraph{When to prefetch?} Prefetching too early or too late could reduce the benefits of prefetching due to potential evictions and overheads. Thus, it is essential to prefetch the required data structure \emph{just-in-time} to maximize the benefits. Therefore, we measure the improvements for different ``spatial prefetching distances'' to find the optimal distance for our use case. We define spatial prefetching distance as the packet gap between the currently-being-processed packet and the upcoming packet that is expected to be processed in the near future.

Next, we will show the potential benefits of just-in-time state prefetching. We only report the throughput of a single-core load balancer due to our workload-generation method that clones packets \& embeds Cuckoo hash values; however, our takeaways could be applicable to multi-core applications and could have a positive impact on latency.


\subsection{Potential Benefits} 

We run a simple experiment with a deterministic packet order and we embed the flow identifier of an upcoming packet into the current packet. We modify the \texttt{FlowIPManagerIMP} element of the load balancer to prefetch the first bucket of its hash table based on the embedded packet identifier before proceeding to the actual lookup operation. Using a trace with deterministic packet order also enables us to investigate the impact of spatial prefetching distance. This helps us to maximize the prefetching benefits, \ie how far in the future (or how many packets ahead) we should see to maximize the throughput improvements.

As upcoming packets are always predictable in our experiments (given their deterministic order), we initially calculate the 5-tuple value of the upcoming (\ie expected-to-be-received) packet based on the currently-being-processed packet's header. This allows us to calculate the Cuckoo hash value of the upcoming packet based on its 5-tuple value.
As expected, this method (``NoOffload'') imposes additional computation overhead on the load balancer; therefore, we also consider an alternative approach (``Offload'') where we calculate the Cuckoo hash of the upcoming packet on a programmable switch and embed the Cuckoo hash of the current and upcoming packets into the current packet. By doing so, we can eliminate the prefetching overhead to a great extent and can see the full potential of network-accelerated just-in-time prefetching.

\vspace{0.5em}

\begin{figure}[hb!]
    \centering
    \includegraphics[width=1\linewidth]{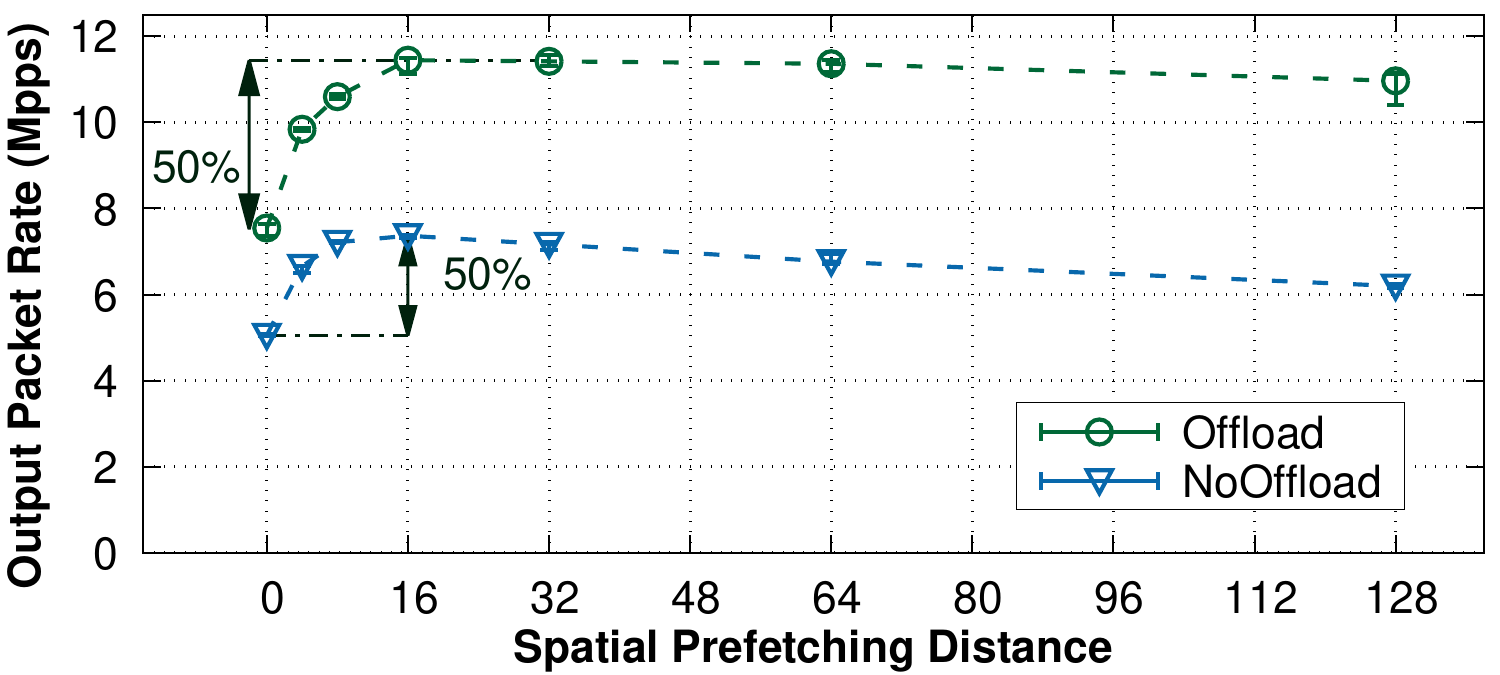}
    \caption{Fine-tuning the spatial prefetching distance is essential to maximize the throughput improvements. 
    }
    \label{fig:eval:flow:prefetch-dist}
\end{figure}

Figure~\ref{fig:eval:flow:prefetch-dist} shows the impact of prefetching states for different spatial prefetching distances. When the spatial prefetching distance is small, the prefetched data will become available in caches \emph{too late}; therefore, the load balancer does not experience the full potential of the state prefetching. On the other hand, large spatial prefetching distances load the data \emph{too early} into the cache, which increases the probability of the data/state being evicted from the cache before the (expected) packet arrives. This experiment demonstrates that performing \emph{just-in-time} prefetching is crucial to maximizing the benefits of state prefetching. Moreover, these results show that the maximum benefit occurs when prefetching distance is 16 with and without hash offloading. Note that DPDK typically receives 32 or fewer packets in each batch; therefore, the prefetching distance of 16 can benefit both current \& future batches.

\begin{figure}[ht!]
    \centering
    \includegraphics[width=1\linewidth]{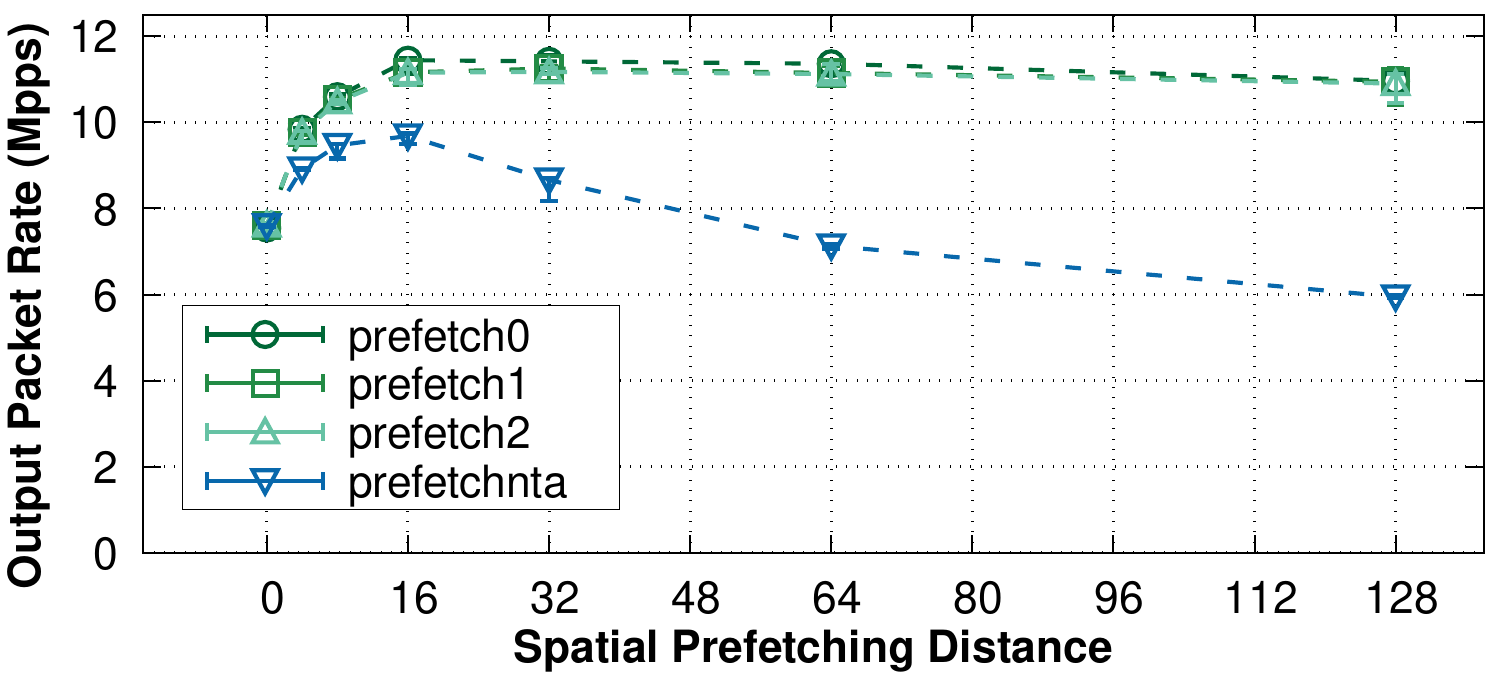}
    \caption{Using \texttt{prefetchnta} reduces the throughput improvements for large spatial prefetching distances, due to its lower temporal \& spatial locality.}
    \label{fig:eval:flow:prefetch-options}
\end{figure}

Figure~\ref{fig:eval:flow:prefetch-options} shows the impact of using different prefetching instructions. These results show that prefetching data to different levels of cache hierarchy has a negligible impact on performance benefits. Additionally, using \texttt{prefetchnta} with the lowest temporal \& spatial factor reduces the throughput improvements for large 
prefetching distances. 

\begin{figure}[h!]
    \centering
    \subfloat[{Throughput.}]{\label{fig:eval:flow:pps}\includegraphics[width=1\linewidth]{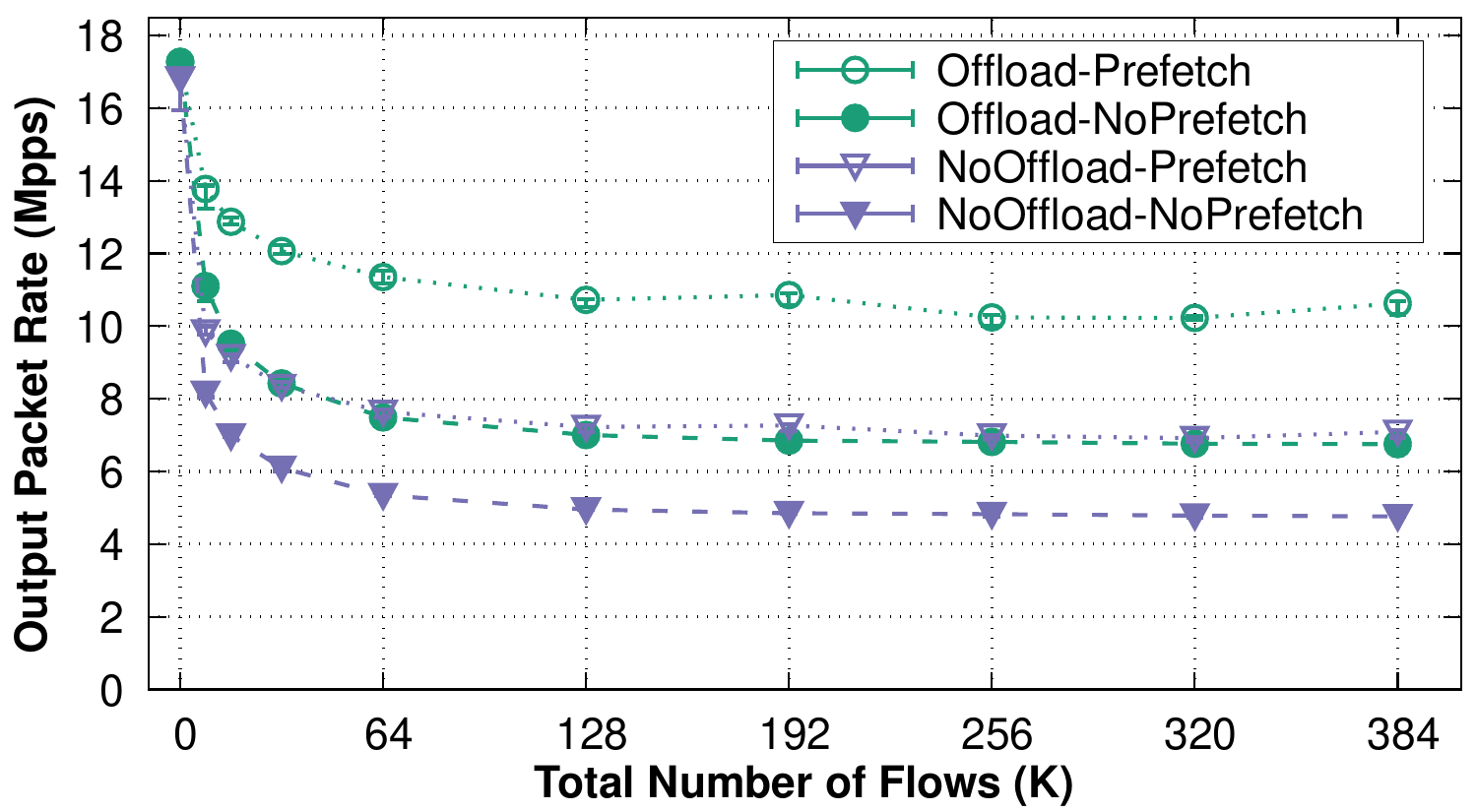}}
    
    \subfloat[{Average number of per-packet LLC misses.}]{\label{fig:eval:flow:llc}\includegraphics[width=1\linewidth]{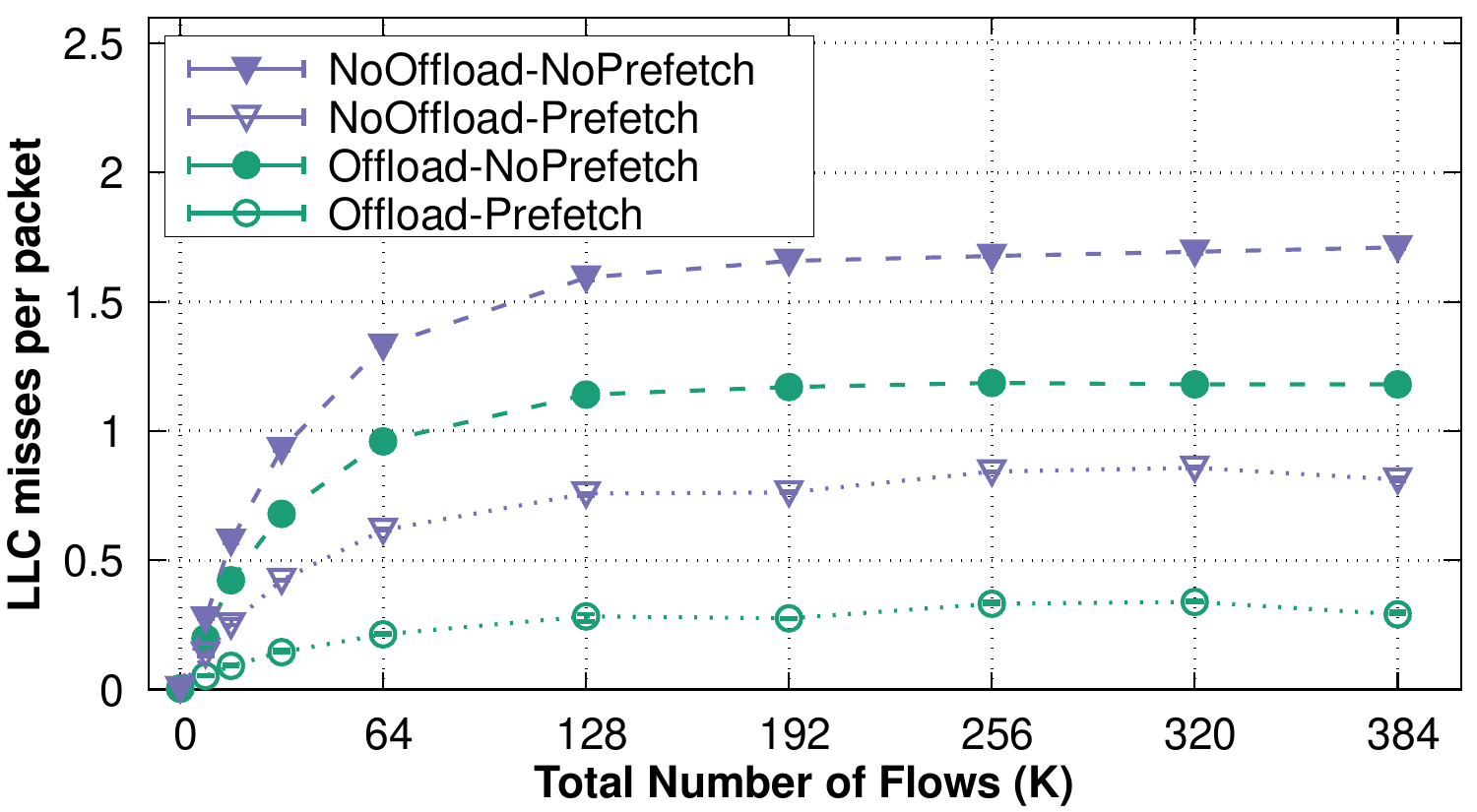}}
    
    \caption{Performing just-in-time prefetching improves the throughput by up to 50\% (\ie it recovers the throughput drop due to statefulness).}
    \label{fig:eval}
\end{figure}

\smartparagraph{Impact on throughput drop.} We conclude our preliminary analysis by demonstrating the impact of carefully-timed prefetching on throughput and the number of per-packet LLC misses when an L4 load balancer is receiving different numbers of flows. Figure~\ref{fig:eval} shows that performing prefetching improves the throughput by up to 50\% (\ie it recovers the throughput drop due to statefulness); similarly, it reduces the number of per-packet LLC misses. As expected, when we offload the hash calculation to the programmable switch, the load balancer achieves higher throughput and experiences a smaller number of LLC misses. Furthermore, performing prefetching (without offloading) results in similar throughput to the no-prefetching case (with offloading) despite spending some cycles on hash calculation. 

\section{Building a Just-in-time Prefetcher}
\label{sect:approaches}

The last section showed the potential benefits of just-in-time state prefetching for stateful networking applications. However, deploying a real-world solution requires knowing the estimated arrival time of upcoming packets and performing the prefetching as efficiently as possible. 

This section explores different alternatives for building a just-in-time prefetcher, called \systemname, which is responsible for \first providing prefetching hints to backend servers and \Second prefetching the required data structures. 

\smartparagraph{(i) How to provide hints?} The first step is to estimate the arrival time of upcoming packets. This hinting can potentially be implemented in different entities in the network that inform the backend servers regarding the time and contents to be prefetched. We envision two alternatives for providing the prefetching hints.
\begin{itemize}[leftmargin=*,noitemsep]
    \item  Clients can incorporate signals to specify an estimate to the backend servers regarding the arrival time of the next packet in a flow. This approach can potentially be accompanied by a mechanism to further fine-tune the estimated time. For instance, a backend server can track the number of per-packet cache misses and use a control loop to more timely prefetch the required data structures. Moreover, some applications send traffic periodically with a known interval (\eg video streaming), this interval can be used by stateful applications as a prefetching signal. When embedding such information, security measures should be considered to prevent potential attacks.
    \item A programmable network device (\eg a programmable switch or a smart NIC) can send a signal to the backend servers regarding the arrival time of the upcoming packets. The device can either rely on (1) buffering packets in the network or (2) employing statistical pattern recognition techniques (\eg AI-based algorithms) to estimate the arrival time and provide hints to the backend servers. The former method can be already implemented in existing programmable switches, as they currently support packet recirculation, making it possible to delay/buffer packets in the network to enforce a deterministic packet order.
\end{itemize}

\smartparagraph{(ii) How to prefetch?} After crafting the prefetching hints, the next step is to efficiently use the received information. As shown in \S\ref{sect:motivation:perf}, an obvious method to use the prefetching hints is to modify the stateful applications to perform the prefetching at an appropriate time. This method would potentially provide the most locality for the prefetched data, as the required data structures would be loaded into the highest level (L1) cache of the core responsible for processing the upcoming packet. However, it may not be preferable in some scenarios, since it requires modifying the application's code and executing additional instructions to prefetch the data. An alternative implementation can exploit other available resources to perform the prefetching. For example, a smart NIC can be configured to \first extract the prefetching information and \Second redirect them to cores specifically deployed to handle prefetching requests, or \Secondb to perform the prefetching itself. 
Step \Second prefetches the data into other cores' caches, which requires inter-core communication to move the states to the processing core; therefore, this method may be less efficient than performing prefetching by the application. However, it is still beneficial, as it mitigates loading states from the memory \emph{and} eliminates the need to modify the application's code. It is worth mentioning that step \first can be potentially omitted if the clients and/or network devices send prefetching hints as standalone packets rather than incorporating the information into the existing packets; however, this alternative would consume more network bandwidth.

\section{Future Use Cases and Directions}
\label{sect:usecases}

This section elaborates on \first future use cases of state prefetching and \Second further optimizations.


 
 \subsection{Applications}

\smartparagraph{Network functions.} This paper primarily focused on an L4 load balancer; however, just-in-time prefetching can be beneficial for \emph{other} stateful network functions, such as advanced packet schedulers or a TCP optimizer.

\smartparagraph{Congestion control protocols.} Many networking applications rely on either Linux-based or userspace congestion control protocols (\eg TCP and QUIC) to ensure reliable \& fair transmission. These protocols operate at the granularity of a flow, which requires them to keep per-flow states; therefore, a highly optimized implementation of congestion control protocol could potentially benefit from just-in-time state prefetching to optimize the protocol stack processing.

\smartparagraph{Key-value stores.} Key-value stores are one of the main building blocks of Internet services. They often act as an intermediate cache layer between slow database servers and frontend servers (\eg web servers) to hide the additional latency of storage systems. Similar to stateful load balancers, key-value stores mostly employ hash tables to store the key-value pairs; therefore, prefetching the key-value pairs in advance could improve their performance. However, providing hints to the servers requires a different traffic pattern recognition since key-value GET/SET requests are not necessarily flow dependant, \ie each flow can hypothetically access all key-value pairs. Key-value store workloads are often skewed, \ie some keys are requested more than others; hence, popular key-value pairs may have a higher probability of being available in the cache; however, less popular large key-value pairs could evict the popular ones from the cache, making just-in-time prefetching beneficial.

 \subsection{Programmable Hardware \& Accelerators}
 
 We focused on commodity CPU-based hardware with cache memories; however, other networking equipment with a hierarchical memory can benefit from our solution. For instance, a recent wave of publications~\cite{tea,dart,switchml,ribosome} extends the limited memory on programmable switches with disaggregated memory accessible from RDMA-capable servers to address the challenges of implementing advanced networking applications. In those proposals, a programmable switch may need to fetch some data structures from the remote memory, where performing in-advance prefetching could hide the imposed fetching/loading latency and improve performance. Moreover, our solution could be applicable to \first disaggregated servers to minimize remote memory access latency~\cite{far-memory,mind-dissagg,SDHI-amir} and \Second smart NICs to hide PCIe overhead when using the host memory~\cite{nicbench-katsikas,understand_pcie}.

 \subsection{Optimizing Data Structures and Code}

This paper mainly considered the benefits of just-in-time prefetching for a two-layer implementation of Cuckoo hashing in DPDK; however, any other hash table or data structure could potentially benefit from our approach. For instance, one can potentially embed the data in the hash table (rather than the index of an array), thus prefetching the actual data without imposing computational overhead. An implementation of \systemname could potentially parse a DPDK program (along with a workload profile similar to PGO) and automatically include prefetching instructions and/or \textit{optimize data structures} for the performance-critical data structures keeping per-flow states. For instance, if a programmable device guarantees a deterministic order for the packets, \systemname can replace a hash table with a queue to mitigate unnecessary overheads of keeping states.

 \subsection{Further Cache Optimizations}
 
 This paper only considered using \emph{prefetching} instructions to efficiently handle packet states for networking applications. Future processors are expected to be shipped with more cache management instructions \& techniques. For instance, Intel Xeon ``Sapphire Rapids'' processors are going to feature a ``\texttt{CLDEMOTE}'' instruction that can be used to demote cache lines by moving them to lower/further levels in the cache hierarchy. One can exploit this instruction to better manage the cache for stateful applications based on prefetching hints.

\section{Conclusion}
\label{sect:conclusions}

Statefulness can make cache memories less effective for high-speed networking applications. This paper proposes an unexplored path to notify the applications in advance about upcoming packets, enabling them to prefetch the data structures required for packet processing into the cache \emph{before} the arrival of the actual packets. Our goal is to emphasize the importance of exploring new opportunities and developing modern techniques to better take advantage of cache memories at multi-100-Gbps rates.

\bibliographystyle{abbrv}
\begin{small}
\bibliography{ref}

\begin{thebibliography}{10}

\bibitem{far-memory}
E.~Amaro, C.~Branner-Augmon, Z.~Luo, A.~Ousterhout, M.~K. Aguilera, A.~Panda,
  S.~Ratnasamy, and S.~Shenker.
\newblock {Can Far Memory Improve Job Throughput?}
\newblock In {\em Proceedings of the Fifteenth European Conference on Computer
  Systems}, EuroSys '20, New York, NY, USA, 2020. Association for Computing
  Machinery.

\bibitem{fastclick}
T.~Barbette, C.~Soldani, and L.~Mathy.
\newblock {Fast Userspace Packet Processing}.
\newblock In {\em Proceedings of the Eleventh ACM/IEEE Symposium on
  Architectures for Networking and Communications Systems}, ANCS '15, pages
  5--16, Washington, DC, USA, 2015. IEEE Computer Society.

\bibitem{cheetah-lb}
T.~Barbette, C.~Tang, H.~Yao, D.~Kosti{\'c}, G.~Q. {Maguire Jr.},
  P.~Papadimitratos, and M.~Chiesa.
\newblock {A High-Speed Load-Balancer Design with Guaranteed
  Per-Connection-Consistency }.
\newblock In {\em 17th {USENIX} Symposium on Networked Systems Design and
  Implementation ({NSDI} 20)}, pages 667--683, Santa Clara, CA, Feb. 2020.
  {USENIX} Association.

\bibitem{hXDP}
M.~S. Brunella, G.~Belocchi, M.~Bonola, S.~Pontarelli, G.~Siracusano,
  G.~Bianchi, A.~Cammarano, A.~Palumbo, L.~Petrucci, and R.~Bifulco.
\newblock {hXDP}: Efficient software packet processing on {FPGA} {NICs}.
\newblock In {\em 14th USENIX Symposium on Operating Systems Design and
  Implementation (OSDI 20)}, pages 973--990. USENIX Association, Nov. 2020.

\bibitem{bbr}
N.~Cardwell, Y.~Cheng, S.~H. Yeganeh, I.~Swett, and V.~Jacobson.
\newblock {BBR Congestion Control}.
\newblock Internet-Draft draft-cardwell-iccrg-bbr-congestion-control-02,
  Internet Engineering Task Force, Mar. 2022.
\newblock Work in Progress.

\bibitem{lb-offload}
T.~Cui, W.~Zhang, K.~Zhang, and A.~Krishnamurthy.
\newblock {\em {Offloading Load Balancers onto SmartNICs}}, page 56–62.
\newblock Association for Computing Machinery, New York, NY, USA, 2021.

\bibitem{redundant_elimination}
B.~Deng, W.~Wu, and L.~Song.
\newblock {Redundant Logic Elimination in Network Functions}.
\newblock In {\em Proceedings of the Symposium on SDN Research}, SOSR ’20,
  page 34–40, New York, NY, USA, 2020. Association for Computing Machinery.

\bibitem{cuckooExtension-2007balanced}
M.~Dietzfelbinger and C.~Weidling.
\newblock {Balanced allocation and dictionaries with tightly packed constant
  size bins}.
\newblock {\em Theoretical Computer Science}, 380(1-2):47--68, 2007.

\bibitem{farshin-packetmill}
A.~Farshin, T.~Barbette, A.~Roozbeh, G.~Q. {Maguire Jr.}, and D.~Kosti\'{c}.
\newblock {PacketMill: Toward per-Core 100-Gbps Networking}.
\newblock ASPLOS 2021, page 1–17, New York, NY, USA, 2021. Association for
  Computing Machinery.

\bibitem{farshin-slice-aware}
A.~Farshin, A.~Roozbeh, G.~Q. {Maguire Jr.}, and D.~Kosti\'{c}.
\newblock {Make the Most out of Last Level Cache in Intel Processors}.
\newblock In {\em Proceedings of the Fourteenth EuroSys Conference 2019},
  EuroSys '19, pages 8:1--8:17, New York, NY, USA, 2019. ACM.

\bibitem{farshin-ddio}
A.~Farshin, A.~Roozbeh, G.~Q. {Maguire Jr.}, and D.~Kosti\'{c}.
\newblock {Reexamining Direct Cache Access to Optimize I/O Intensive
  Applications for Multi-hundred-gigabit Networks}.
\newblock In {\em 2020 {USENIX} Annual Technical Conference ({USENIX} {ATC}
  20)}, pages 673--689. {USENIX} Association, July 2020.

\bibitem{om-hamid}
H.~Ghasemirahni, T.~Barbette, G.~P. Katsikas, A.~Farshin, A.~Roozbeh,
  M.~Girondi, M.~Chiesa, G.~Q. {Maguire Jr.}, and D.~Kosti{\'c}.
\newblock {Packet Order Matters! Improving Application Performance by
  Deliberately Delaying Packets}.
\newblock In {\em 19th USENIX Symposium on Networked Systems Design and
  Implementation (NSDI 22)}, pages 807--827, Renton, WA, Apr. 2022. USENIX
  Association.

\bibitem{flowmage}
H.~Ghasemirahni, A.~Farshin, M.~Scazzariello, M.~Chiesa, and D.~Kosti\'{c}.
\newblock Deploying stateful network functions efficiently using large language
  models.
\newblock In {\em Proceedings of the 4th Workshop on Machine Learning and
  Systems}, EuroMLSys '24, page 28–38, New York, NY, USA, 2024. Association
  for Computing Machinery.

\bibitem{fajita}
H.~Ghasemirahni, A.~Farshin, M.~Scazzariello, G.~Q. Maguire~Jr., D.~Kosti\'{c},
  and M.~Chiesa.
\newblock Fajita: Stateful packet processing at 100 million pps.
\newblock {\em Proc. ACM Netw.}, 2(CoNEXT3), September 2024.

\bibitem{massimo-connection}
M.~Girondi, M.~Chiesa, and T.~Barbette.
\newblock {High-speed Connection Tracking in Modern Servers}.
\newblock In {\em 2021 IEEE 22nd International Conference on High Performance
  Switching and Routing (HPSR)}, pages 1--8, 2021.

\bibitem{gro-dpdk}
J.~Hu.
\newblock {Accelerating Packet Processing with GRO and GSO in DPDK}, 2018.
\newblock
  https://www.dpdk.org/wp-content/uploads/sites/35/2018/06/GRO-GSO-Libraries-Bring-Significant-Performance-Gains-to-DPDK-based-Applications.pdf.

\bibitem{tofino}
{Intel}.
\newblock {Intel Tofino Series}, 2022.
\newblock
  \url{https://www.intel.com/content/www/us/en/products/network-io/programmable-ethernet-switch/tofino-series.html}.

\bibitem{intel-prefetch}
{Intel}.
\newblock {MM\_PREFETCH - Development Reference Guides}, 2022.
\newblock
  \url{https://www.intel.com/content/www/us/en/develop/documentation/fortran-compiler-oneapi-dev-guide-and-reference/top/language-reference/a-to-z-reference/m-to-n/mm-prefetch.html}.

\bibitem{nicbench-katsikas}
G.~P. Katsikas, T.~Barbette, M.~Chiesa, D.~Kosti{\'{c}}, and G.~Q. {Maguire
  Jr.}
\newblock {What You Need to Know About (Smart) Network Interface Cards}.
\newblock In O.~Hohlfeld, A.~Lutu, and D.~Levin, editors, {\em Passive and
  Active Measurement}, pages 319--336, Cham, 2021. Springer International
  Publishing.

\bibitem{metron}
G.~P. Katsikas, T.~Barbette, D.~Kosti{\'c}, R.~Steinert, and G.~Q. {Maguire
  Jr.}
\newblock {Metron: {NFV} Service Chains at the True Speed of the Underlying
  Hardware}.
\newblock In {\em 15th {USENIX} Symposium on Networked Systems Design and
  Implementation ({NSDI} 18)}, pages 171--186, Renton, WA, 2018. {USENIX}
  Association.

\bibitem{tea}
D.~Kim, Z.~Liu, Y.~Zhu, C.~Kim, J.~Lee, V.~Sekar, and S.~Seshan.
\newblock {TEA: Enabling State-Intensive Network Functions on Programmable
  Switches}.
\newblock In {\em Proceedings of the Annual Conference of the ACM Special
  Interest Group on Data Communication on the Applications, Technologies,
  Architectures, and Protocols for Computer Communication}, SIGCOMM '20, page
  90–106, New York, NY, USA, 2020. Association for Computing Machinery.

\bibitem{dart}
J.~Langlet, R.~Ben-Basat, S.~Ramanathan, G.~Oliaro, M.~Mitzenmacher, M.~Yu, and
  G.~Antichi.
\newblock {Zero-CPU Collection with Direct Telemetry Access}.
\newblock In {\em Proceedings of the Twentieth ACM Workshop on Hot Topics in
  Networks}, HotNets '21, page 108–115, New York, NY, USA, 2021. Association
  for Computing Machinery.

\bibitem{prefetching-journal}
J.~Lee, H.~Kim, and R.~Vuduc.
\newblock {When Prefetching Works, When It Doesn’t, and Why}.
\newblock {\em ACM Trans. Archit. Code Optim.}, 9(1), mar 2012.

\bibitem{mind-dissagg}
S.-s. Lee, Y.~Yu, Y.~Tang, A.~Khandelwal, L.~Zhong, and A.~Bhattacharjee.
\newblock {MIND: In-Network Memory Management for Disaggregated Data Centers}.
\newblock In {\em Proceedings of the ACM SIGOPS 28th Symposium on Operating
  Systems Principles}, SOSP '21, page 488–504, New York, NY, USA, 2021.
  Association for Computing Machinery.

\bibitem{batchy}
T.~L{\'e}vai, F.~N{\'e}meth, B.~Raghavan, and G.~Retvari.
\newblock {Batchy: Batch-scheduling Data Flow Graphs with Service-level
  Objectives }.
\newblock In {\em 17th {USENIX} Symposium on Networked Systems Design and
  Implementation ({NSDI} 20)}, pages 633--649, Santa Clara, CA, Feb. 2020.
  {USENIX} Association.

\bibitem{vpp-paper}
L.~Linguaglossa, D.~Rossi, S.~Pontarelli, D.~Barach, D.~Marjon, and P.~Pfister.
\newblock {High-speed data plane and network functions virtualization by
  vectorizing packet processing}.
\newblock {\em Computer Networks}, 149:187--199, 2019.
\newblock
  \url{https://www.sciencedirect.com/science/article/pii/S1389128618312957}.

\bibitem{contention-aware-justine}
A.~Manousis, R.~A. Sharma, V.~Sekar, and J.~Sherry.
\newblock {Contention-Aware Performance Prediction For Virtualized Network
  Functions}.
\newblock In {\em Proceedings of the Annual Conference of the ACM Special
  Interest Group on Data Communication on the Applications, Technologies,
  Architectures, and Protocols for Computer Communication}, SIGCOMM '20, page
  270–282, New York, NY, USA, 2020. Association for Computing Machinery.

\bibitem{morpheus}
S.~Miano, A.~Sanaee, F.~Risso, G.~R\'{e}tv\'{a}ri, and G.~Antichi.
\newblock {Domain Specific Run Time Optimization for Software Data Planes}.
\newblock In {\em Proceedings of the 27th ACM International Conference on
  Architectural Support for Programming Languages and Operating Systems},
  ASPLOS 2022, page 1148–1164, New York, NY, USA, 2022. Association for
  Computing Machinery.

\bibitem{ericsson-switch}
L.~Moln\'{a}r, G.~Pongr\'{a}cz, G.~Enyedi, Z.~L. Kis, L.~Csikor, F.~Juh\'{a}sz,
  A.~K\H{o}r\"{o}si, and G.~R\'{e}tv\'{a}ri.
\newblock {Dataplane Specialization for High-Performance OpenFlow Software
  Switching}.
\newblock In {\em Proceedings of the 2016 ACM SIGCOMM Conference}, SIGCOMM
  ’16, page 539–552, New York, NY, USA, 2016. Association for Computing
  Machinery.

\bibitem{scalable_family}
D.~Mulnix.
\newblock {Intel Xeon Processor Scalable Family Technical Overview}.
\newblock
  \url{https://www.intel.com/content/www/us/en/developer/articles/technical/xeon-processor-scalable-family-technical-overview},
  Sep 2017.
\newblock Online; accessed 2022-06-1.

\bibitem{understand_pcie}
R.~Neugebauer, G.~Antichi, J.~F. Zazo, Y.~Audzevich, S.~L\'{o}pez-Buedo, and
  A.~W. Moore.
\newblock {Understanding PCIe Performance for End Host Networking}.
\newblock In {\em Proceedings of the 2018 Conference of the ACM Special
  Interest Group on Data Communication}, SIGCOMM '18, pages 327--341, New York,
  NY, USA, 2018. ACM.

\bibitem{CAT}
K.~Nguyen.
\newblock {Introduction to Cache Allocation Technology in the Intel Xeon
  Processor E5 v4 Family}, Feb 2016.
\newblock
  {\url{https://www.intel.com/content/www/us/en/developer/articles/technical/introduction-to-cache-allocation-technology.html},
  accessed 2022-06-13}.

\bibitem{connectx5}
{NVIDIA Networking}.
\newblock {NVIDIA Mellanox ConnectX-5 adapters}, 2021.
\newblock \url{https://www.nvidia.com/en-us/networking/ethernet/connectx-5/}.

\bibitem{beamer}
V.~Olteanu, A.~Agache, A.~Voinescu, and C.~Raiciu.
\newblock {Stateless Datacenter Load-balancing with Beamer}.
\newblock In {\em 15th USENIX Symposium on Networked Systems Design and
  Implementation (NSDI 18)}, pages 125--139, Renton, WA, Apr. 2018. USENIX
  Association.

\bibitem{cuckoo-hash}
R.~Pagh and F.~F. Rodler.
\newblock {Cuckoo Hashing}.
\newblock {\em J. Algorithms}, 51(2):122–144, may 2004.

\bibitem{SDHI-amir}
A.~Roozbeh, J.~Soares, G.~Q. {Maguire Jr.}, F.~Wuhib, C.~Padala, M.~Mahloo,
  D.~Turull, V.~Yadhav, and D.~Kosti{\'c}.
\newblock {Software-Defined ``Hardware'' Infrastructures: A Survey on Enabling
  Technologies and Open Research Directions}.
\newblock {\em IEEE Communications Surveys Tutorials}, 20(3):2454--2485,
  thirdquarter 2018.

\bibitem{switchml}
A.~Sapio, M.~Canini, C.-Y. Ho, J.~Nelson, P.~Kalnis, C.~Kim, A.~Krishnamurthy,
  M.~Moshref, D.~Ports, and P.~Richtarik.
\newblock {Scaling Distributed Machine Learning with {In-Network} Aggregation}.
\newblock In {\em 18th USENIX Symposium on Networked Systems Design and
  Implementation (NSDI 21)}, pages 785--808. USENIX Association, Apr. 2021.

\bibitem{ribosome}
M.~Scazzariello, T.~Caiazzi, H.~Ghasemirahni, T.~Barbette, D.~Kosti{\'c}, and
  M.~Chiesa.
\newblock A {High-Speed} stateful packet processing approach for tbps
  programmable switches.
\newblock In {\em 20th USENIX Symposium on Networked Systems Design and
  Implementation (NSDI 23)}, pages 1237--1255, Boston, MA, Apr. 2023. USENIX
  Association.
\newblock
  \url{https://www.usenix.org/conference/nsdi23/presentation/scazzariello}.

\bibitem{RESQ-scheduling}
A.~Tootoonchian, A.~Panda, C.~Lan, M.~Walls, K.~Argyraki, S.~Ratnasamy, and
  S.~Shenker.
\newblock {ResQ: Enabling SLOs in Network Function Virtualization}.
\newblock In {\em 15th {USENIX} Symposium on Networked Systems Design and
  Implementation ({NSDI} 18)}, pages 283--297, Renton, WA, Apr. 2018. {USENIX}
  Association.

\bibitem{pegasus}
Z.~Zhao, H.~Sadok, N.~Atre, J.~C. Hoe, V.~Sekar, and J.~Sherry.
\newblock {Achieving 100Gbps Intrusion Prevention on a Single Server}.
\newblock In {\em 14th {USENIX} Symposium on Operating Systems Design and
  Implementation ({OSDI} 20)}, pages 1083--1100. {USENIX} Association, Nov.
  2020.

\end{thebibliography}
\end{small}

\end{document}